# Fermi-level depinning in metal–2D multilayered semiconductor junctions

Qian Wang,[1,2] Yangfan Shao,[1,3] Penglai Gong,[1] and Xingqiang Shi[1,a]

[1] Department of Physics and Guangdong Provincial Key Laboratory for Computational Science and Material Design, Southern University of Science and Technology, Shenzhen 518055, China

[2] Harbin Institute of Technology, Harbin 150080, China

[3] Joint Key Laboratory of the Ministry of Education, Institute of Applied Physics and Materials Engineering, University of Macau, Macau, China

[a] E-mail: shixq@sustech.edu.cn

## Abstract

Thicknesses-dependent performance of metal–two-dimensional (2D) semiconductor junctions (MSJ) in electronics/optoelectronics have attracted increasing attention, but till present, people have little knowledge about the micro-mechanism of the thicknesses (or layer-number) dependence. Here, by first-principles calculations based on density functional theory, we show that the Fermi-level pinning (FLP) factor of MSJ depends sensitively on the layer-number of few-layer 2D semiconductors, and, an extended FLP theory is proposed for metal–2D *multilayered* semiconductor junctions (M*m*SJ). Taking multilayered $MoS_2$ as a typical example for van der Waals (vdW) semiconductor in M*m*SJ, the extended FLP theory has the following character: strong pinning *right* at the metal–1$^{st}$-layer semiconductor interface while depinning occurs between $MoS_2$ layers. This depinning effect between vdW layers has several important consequences: 1) the overall pinning in M*m*SJ is greatly weakened; 2) *p*-type contact, rarely obtained in metal–monolayer $MoS_2$ junctions, becomes favored in M*m*SJ, which is important for CMOS logical devices; 3) depinning between $MoS_2$ layers result in type II band alignment in $MoS_2$ 'homojunction' supported on metals, which is useful for optoelectronics. Moreover, our extended FLP theory sheds light on the recent controversial experimental observation and paves a new and universal route to type II band alignment in vdW 'homojunctions'.

# 1. Introduction

Semiconductors have been at the heart of electronic/optoelectronic devices over decades. As a typical example for van der Waals (vdW) semiconductor, two dimensional (2D) layered $MoS_2$, which possesses sizable band gap [1,2], high on/off current ratio [3] and high room-temperature carrier mobility [4-6], has proved highly promising for the construction of electronic/optoelectronic devices. However, a bottleneck problem in 2D $MoS_2$ (and other 2D semiconductors) device applications is the unexpectedly high electrical contact resistance resulted from Fermi-level pinning (FLP) in three-dimension (3D) metal–2D $MoS_2$ junctions [7,8]. Although different interface engineering ways or using 2D metal are proposed to reduce FLP [9-11], the relative simple traditional 3D metal-2D semiconductor junction without further engineering (such as insertion an insulator layer) are widely concerned since the simple process, easy operation and stable structures [12]. The most important parameter for contact resistance in a metal-semiconductor junction (MSJ) is the Schottky barrier height (SBH), which can fundamentally determine charge transport efficiency and affect device performance [13]. Unfortunately, the knowledge about the micro-mechanism of SBH at MSJ and the related FLP theory is still incomprehensive and needs to be elucidated.

The mechanism of SBH and pinning factor $S$ (which describes the strength of FLP and the difficulty of tuning SBH by changing metal work-functions) [14] in metal–*monolayer* $MoS_2$ heterojunctions have been studied intensively [3,15-18]. However, almost no one is aware of the impact of $MoS_2$ thickness (or layer-number) on pinning factor $S$, which subsequently affected SBH. Although, thicknesses-dependent performance in metal–*multilayered* semiconductor junctions (M*m*SJ) have attracted increasing attention [12,19-28], clear explanations about layer-number-dependent FLP mechanism and pinning factor $S$ have not been reported till present. Theories derived from 3D metal-monolayer $MoS_2$ junctions (MSJ) are used to explain the layer-number-dependent performance in (M*m*SJ), which may lead to confusing understandings.

To solve the above problem, in the current work, an extended FLP theory is proposed for M*m*SJ, namely, the Schottky barrier height/type and the pinning factor $S$ is layer-number dependent. From literature, strong pinning ($S \cong 0.3$) in 3D metal-monolayer $MoS_2$ junctions (with ideal interface)[4,17,27] is widely accepted; while very recently, $S \cong 1$ for 3D metal-*multilayered* $MoS_2$ junctions is reported in experiment [12], which means that the Schottky–Mott limit without pinning is obtained. The reason why pinning factor $S$ undergoes such a huge transformation and why the FLP goes from strong for monolayer to weak for multilayer is the depinning effect between the van der Waals $MoS_2$ layers, which is summarized below.

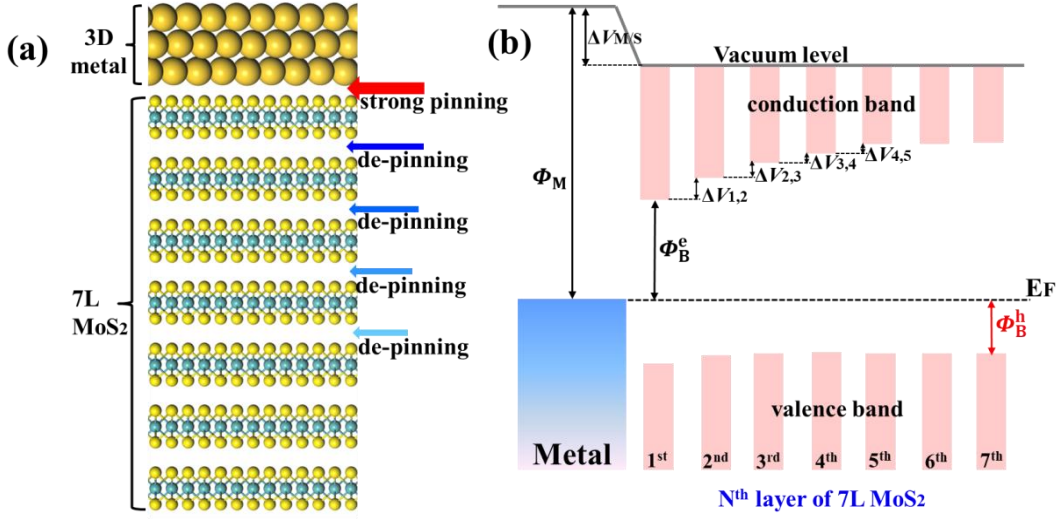

**FIG. 1. Strong pinning right at the metal–1$^{st}$-layer $MoS_2$ interface but depinning between $MoS_2$ layers.** (a) Sketch maps of depinning effect in metal-7L $MoS_2$. (b) Band alignment in a metal–7L $MoS_2$ junction. $\Phi_B^e/\Phi_B^h$ is the Schottky barrier height for electrons/holes, respectively. $\Phi_M$ is the work functions of clean metal surface. $\Delta V_{M/S}$ is the potential step at the metal–$MoS_2$ interface, and $\Delta V_{n-1,n}$ (subscript $n$ denotes the $n^{th}$ $MoS_2$ layer) is the band offsets between $MoS_2$ layers.

The strong pinning right at the metal-$MoS_2$ interface and depinning between the van der Waals layers of $MoS_2$ are sketched in Figure 1. In Figure 1b, the band alignment of each layer in the metal–7L $MoS_2$ junction is shown. The first $MoS_2$ layer is strongly pinned due to metal-induced-gap-states and interface dipole [4,29]. In the subsequent $MoS_2$ layers away from the metal surface, the pinning effect is gradually removed due to a depinning effect occurs between $MoS_2$ layers, which results in the band offsets $\Delta V_{n-1,n}$ in conduction band minimum (CBM) between $MoS_2$ layers. For valance band maximum (VBM), the band offset is not so obvious due to the indirect band character of multilayer $MoS_2$ (see last paragraph in part B in Section II). The depinning effect decays as the $MoS_2$ layer far away from the metal surface (decays to zero at about the 5$^{th}$ $MoS_2$ layer from our first-principles calculation, as sketched in Figure 1a & 1b). The depinning effect between $MoS_2$ layers results in the overall weakened pinning in M$m$SJ. Detailed discussions are given in the Section below.

## 2. Results and Discussion

### A. Layer-number dependent pinning factor $S$ and $n$- to $p$-type contact transition

The pinning factor, $S$, which describes the strength of FLP, is defined as the slope of the $\Phi_B$-versus-$\Phi_M$ for a given semiconductor with a set of metals:

$$S = \left|\frac{d\Phi_B}{d\Phi_M}\right| \qquad (1),$$

where $\Phi_B$ is the SBH and $\Phi_M$ is the metal work function. As defined, $S$ characterizes the difficulty of changing SBH by metal work function. Two extremums, 1 and 0, of $S$ represent no pinning corresponding to Schottky-Mott limit [30] and the strong pinning (Schottky barrier does not change with the metal work function) corresponding to Bardeen limit [31], respectively. In Schottky-Mott limit, the SBH $\Phi_B$ is obtained by band alignment of the non-interacting subsystems:

$$\Phi_B^e = \Phi_M - EA \; ; \quad \Phi_B^h = IP - \Phi_M \qquad (2),$$

where $\Phi_B^e$ and $\Phi_B^h$ are the SBH for electrons (n-type) and holes (p-type), $EA$ and $IP$ are the electron affinity and ionization potential of the semiconductor. Considering FLP in MSJ, $\Phi_B$ is determined by the energy difference between the Fermi level and the semiconductor band edges in the junction (as shown in Figure 1b and Figure 3b):

$$\Phi_B^e = E_{CBM} - E_F \; ; \quad \Phi_B^h = E_F - E_{VBM} \qquad (3),$$

where $E_F$ is the Fermi level, and $E_{VBM}$ and $E_{CBM}$ are VBM and CBM of the projected band structure of semiconductor in the MSJ or M$m$SJ. In Eq. 3, the pinning effect is automatically included in first-principles self-consistent calculations.

For *monolayer* MoS$_2$ contacts with 3D metals, strong pinning at the metal-semiconductor (M-S) interface is formed. The FLP mainly originates from 1) the formation of interface electrical dipole due to charge transfer/redistribution at the M-S interface and 2) gap states (GS) from different origins, including metal induced gap states (MIGS) [32], interaction induced gap states (IIGS) [4], and defect/disorder induced gap states (DIGS) [16]. The DIGS can be neglected for high-quality ideal interface. The existence of MIGS/IIGS and interface dipole pins Fermi energy to a certain energy level, making $\Phi_B$ deviate from that in the Schottky-Mott limit. As a result, the pinning factor $S$ gets a value between 0 and 1. For monolayer MoS$_2$, the pinning factor $S$ achieved two typical values in literature, 1) $S \cong 0.3$ [4,27] resulted from high quality metal–MoS$_2$ contacts without chemical disorder/defects at the M-S interface, and 2) $S \cong 0.1$ [19,33] obtained from MoS$_2$ with deposited/evaporated electrodes, which result in considerable defect states in the MoS$_2$ band gap and finally lead to a very stronger FLP [34]. In summary, even with an ideal metal–MoS$_2$ interface, strong pinning with a pinning factor of $S \cong 0.3$ is often obtained.

However, very recently, almost no pinning with a pinning factor $S = 0.96$ was reported in experiment [12] for *multilayered* MoS$_2$ contacts with 3D metals. Considering the GS (include MIGS and IIGS) and interface dipole at ideal M-S interface, no pinning results can only be obtained in M*m*SJ with *multilayered* MoS$_2$ but not MSJ with single layer MoS$_2$. So, the mechanisms of depinning effect between layers of multilayered MoS$_2$ must be taken into account, which go well beyond the FLP mechanism of monolayer MoS$_2$. In the following, we probe the layer-number dependent pinning factor $S$ and its origin.

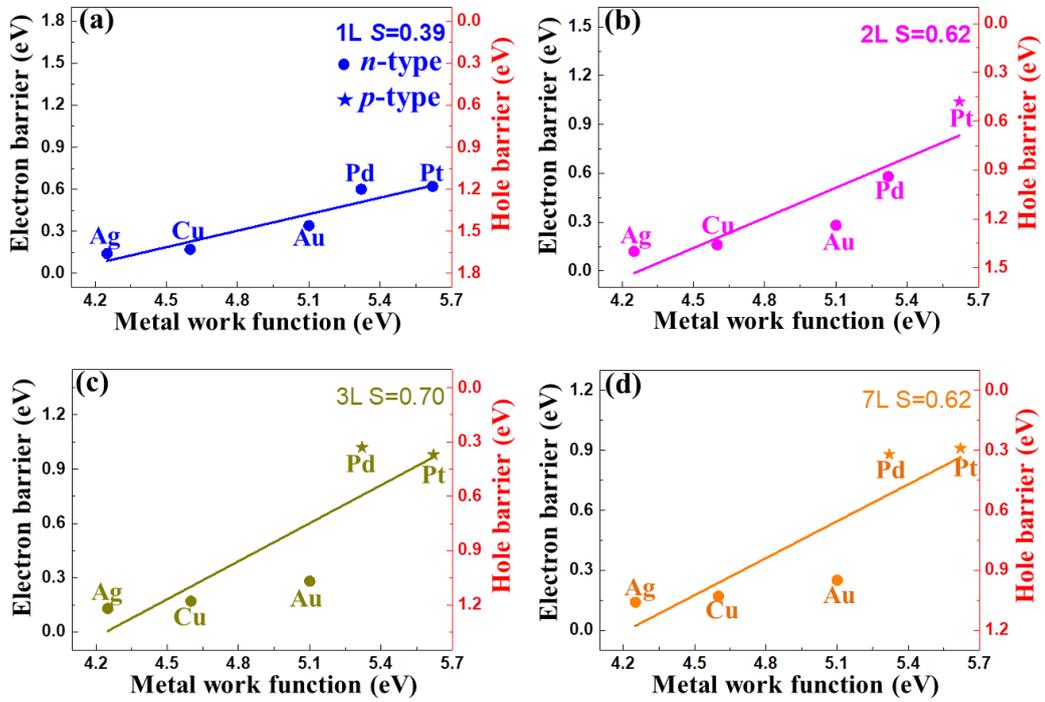

**FIG. 2. Layer-number dependent pinning factor $S$ in M*m*SJ.** (a-d) Schottky barrier height and pinning factor $S$ for M*m*SJ with 1L, 2L, 3L and 7L MoS$_2$. Dots represent *n*-type Schottky barriers, and stars represent *p*-type

Schottky barriers. For junctions with multilayered $MoS_2$, $S$ increase a lot compared to the monolayer (1L) case.

SBHs of metal–$MoS_2$ junctions with different number of layers of $MoS_2$ are shown in Figure 2. The SBHs is calculated by Eq. 3, which is derived from the band structure in M$m$SJ projected to $MoS_2$ (See Figure 3b and Figures S2-4 and Table SIII in Supporting Information). From the SBHs, the Schottky pinning factors $S$ are fitted by Eq. 1. The pinning factors $S$ are 0.39, 0.62, 0.70 and 0.62 for 1L, 2L, 3L and 7L $MoS_2$, respectively. The $S = 0.39$ for 1L $MoS_2$ is well below unity and implies that FLP at the M-S interface is strong. Using slab model with a smaller lattice mismatch for metal-1L $MoS_2$ heterojunction, a pinning factor $S = 0.26$ was obtained (see calculation details in Section IV and discussions in Part V of Supporting Information). The $S$ values are around 0.3, which is in consistent with previous theoretical results [4,15]. From monolayer to bilayer, a notable increase of $S$ to 0.62 is found and the maximum $S$ of 0.70 appears for trilayer $MoS_2$. The $S = 0.70$ with trilayer $MoS_2$ is much larger than that with monolayer $MoS_2$, and is closer to the Schottky-Mott limit. For metal–1L $MoS_2$ junctions in Figure 2a, all of the Schottky barriers (SBs) are *n*-type. In metal–*multilayered* $MoS_2$ junctions, *p*-type SBs appears for metals with larger work functions, such as Pt–2L $MoS_2$ in Figure 2(b). When the $MoS_2$ layer-number is greater than 2, junctions with both Pt and Pd show *p*-type contacts (Figure 2c and 2d). The appearance of *p*-type contacts in Pt–7L $MoS_2$ and Pd–7L $MoS_2$ junctions is in good agreement with experiment [12].

The larger $S$ in junctions with multilayer $MoS_2$, compared to that with monolayer $MoS_2$, means that the *overall* pinning is weakened. But we found that in metal-*multilayered* $MoS_2$ junctions, the pinning at the metal–1$^{st}$-layer $MoS_2$ interface are not weakened, the weakening of the overall pinning effect should be attributed to the depinning effect between the van der Waals $MoS_2$ layers. In the following, we prove this point through two steps: we 1) present evidence of strong pinning in M$m$SJ with *multilayered* $MoS_2$ and 2) show that this strong pinning occurs only at the metal–1$^{st}$-layer $MoS_2$ interface; and we clarify that depinning occurs between $MoS_2$ layers, which results in the appearance of the weakened overall-pinning.

## B. Interface strong pinning but de-pinning between MoS₂ layers

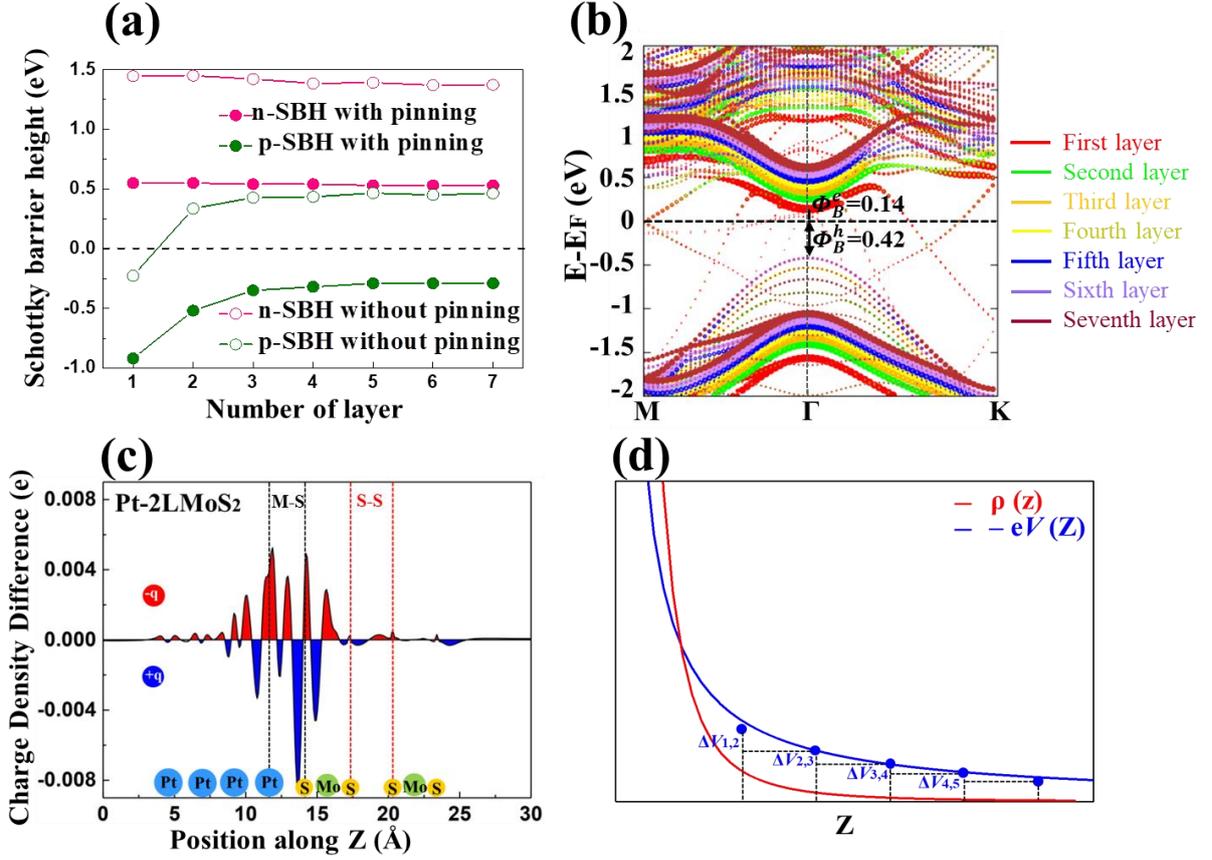

**FIG. 3. Micro-mechanism of pinning at Metal–1st-layer MoS₂ interface and depinning between MoS₂ layers.** (a) Pinning effect M*m*SJ: comparison of SBHs between without pinning (from Schottky–Mott limit) and with pinning (from our first-principles calculation) for 1 to 7 MoS₂ layers, using Pt–*multilayered* MoS₂ contacts as an example. (b) Band structure projected to each layer of MoS₂ in the Ag–7L MoS₂ junction, the contributions from the 1st to 7th layers are marked with different colors. The *n*- and *p*-type SBH are labeled with $\Phi_B^e$ and $\Phi_B^h$, respectively. (c) Electron-density-difference along the vertical z-direction normal to the interface of Pt–2L MoS₂ junction. The metal-semiconductor (M-S) interface is shown with black dotted lines, and the semiconductor-semiconductor (S-S) interface is shown with red dotted lines. (d) MoS₂-layer-averaged charge density difference ρ(z) in the Z-direction and electrostatic potential energy –e$V$(z) with the distance z from the metal surface [35]. The red and blue lines of ρ(z) and –e$V$(z) are schematic diagram based on Eq. 4 below; the blue dots are the first-principles calculation results from a Ag–7L MoS₂ junction.

The fact that the pinning effect exits in metal–*multilayered* MoS₂ junctions can be demonstrated by considering the SBHs with and without pinning, which are shown in Figure 3a. Without pinning, the SBHs satisfy Schottky–Mott limit which is calculated by Eq. 2. With pinning, the SBHs are calculated by Eq. 3, which depends on the projected bands of $MoS_2$ in MSJ. Without pinning, the electron barrier (*n*-SBH) decrease slightly with the increasing of $MoS_2$ layers (the change is within 0.1 eV), while the hole barrier (*p*-SBH) decrease sharply from monolayer to bilayer, make the *p*-SBH from negative to positive (pass through the Fermi level and vanishing *p*-SBH). The sharp decrease is due to the big drop of bandgap, which is mainly due to the direct- to indirect-band gap transition with the number of $MoS_2$ layers change from one to two [36]. With the pinning effect included (Figure 3a), both positive *n*-SBH and negative *p*-SBH appear in the Pt–$MoS_2$ junctions. The *n*-SBH remains at ~0.5 eV with the increasing of $MoS_2$ layer-number, while that in the *without* pinning cases is ~1.4 eV, so there is a 0.9 eV difference for all *n*-SBH between with- and without-pinning cases. Compared with vanishing *p*-SBH in the without pinning case, the appearance of *p*-SBH with pinning demonstrates the existence of pinning in metal–*multilayered* $MoS_2$ junction with the SBH deviation between two cases (with and without pinning). The 0.9 eV difference of *n*-SBH with- and without-pinning also confirms the pinning effect in M*m*SJ.

As discussed before, when monolayer $MoS_2$ contact with 3D metals, strong FLP at metal-$MoS_2$ interface is formed. The origin of FLP is GS and interface dipole. In Ag-7L $MoS_2$ junctions and Pt-2L $MoS_2$ junctions, the effects of GS (Figure 3b, and Figure S3 & S4) and interface dipole (Figure 3c) on the first $MoS_2$ layer are much stronger than that on the other $MoS_2$ layers. As shown in Figure S3 & S4, GS appear in the band gap of $MoS_2$, and this GS are mainly in the first $MoS_2$ layer. There are two sources of GS in our ideal interface system, MIGS and IIGS. In our calculation, the equilibrium separation between Pt (111) surface and the first $MoS_2$ layer is 2.56 Å (the sum of covalent bond radius of platinum and sulfur is 2.42 Å and that of the van der Waals radius is 3.85 Å), suggesting that the interaction between Pt and $MoS_2$ is stronger than typical physical adsorption. The interaction of platinum-sulfur at interface weaken the intralayer sulfur-molybdenum bonding which finally induced the production of GS mainly composed of molybdenum d-orbital characters,[4] this is the so-called IIGS. Apparently, this IIGS can only appear in the $1^{st}$-layer closest to metal surface. Another source is MIGS, which is described as a decay of metal wave function into the semiconductor. The MIGS decay into the first $MoS_2$ layer and the vdW stacking of bilayer $MoS_2$ enhances the decay length of metal wave function [37]. As shown in Figure 3c for Pt-2L $MoS_2$, the charge transfer/redistribution at the metal-MoS$_2$ interface is much larger than that at the MoS$_2$-MoS$_2$ interface. It can be concluded that the interface dipole at metal-MoS$_2$ interface is much larger than at MoS$_2$-MoS$_2$ interface, since interface dipole is caused by charge redistribution at interface. Under the effect of GS and interface dipole, strong pinning appears at metal-1$^{st}$-layer-MoS$_2$ interface.

In Figure 3c, the analysis of differential charge density between the composite junction and the isolated subsystems, i.e. $\Delta\rho = \rho_{Pt/2L\,MoS2} - (\rho_{Pt} + \rho_{2L\,MoS2})$ [4,38], is performed. At the Pt-MoS$_2$ interface, overall, Pt can be considered as gaining electrons while MoS$_2$ lose electrons. There are several mechanisms that affect the charge redistribution [32], which leads to the complex situation as plotted in Figure 3c. The pushback effect played a major role [32]. When MoS$_2$ approaches to Pt surface, the wave functions of the two subsystems overlap [32,39,40]. Pauli repulsion leads to a spatial redistribution of the electron density, in particular to a decrease of the density in the overlap region. Since Pt has reservoir of a large amount of electrons compared to MoS$_2$, the wave functions of Pt are usually more extended and deformable, the net result of this redistribution is that electrons are pushed back into Pt. Therefore, Pt gains electrons and MoS$_2$ loses electrons, thus forming an interface dipole pointing from Pt to MoS$_2$. At the MoS$_2$-MoS$_2$ interface, the dipole direction is opposite to the Pt-MoS$_2$ interface (pointing from the 2$^{nd}$-layer to the 1$^{st}$-layer), which produced a potential step and a depining effect between MoS$_2$ layers in M*m*SJ. As shown in Figure 3c, at the MoS$_2$-MoS$_2$ interface, electrons deplete at the left side and accumulate at the right side, which creates a dipole. The opposite direction of dipole at the MoS$_2$-MoS$_2$ interface origins from the screening effect inside the first MoS$_2$ layer. Note that the 1$^{st}$-layer MoS$_2$ is partial metallized due to GS and hence it can screen the dipole at the metal-MoS$_2$ interface -- indeed an opposite dipole can be seen within the 1$^{st}$-layer of MoS$_2$ from Figure 3c.

Depinning effect is confirmed by the barrier height $\Phi_B^e(n)$ of the $n^{th}$ MoS$_2$ layer, which varies with the layer-number $n$. The $\Phi_B^e(n)$ of each MoS$_2$ layer in Ag-7L MoS$_2$ junctions are listed in Table I. The Schottky barrier in Ag–7L MoS$_2$ junction is *n*-type and the SBH is 0.14 eV which is determined by the first MoS$_2$ layer due to the first-layer contribute the CBM of 7L-MoS$_2$ in Ag-7L MoS$_2$ junctions. There is a gradual upward shift of CBM and increase of $\Phi_B^e(n)$ for the MoS$_2$ layers away from metal surface, and the most obvious increase occurs in the first several layers.

**TABLE I.** The Schottky barrier height $\Phi_B^e(n)$ of the $n^{\text{th}}$ MoS$_2$ layer in the Ag–7L MoS$_2$ junction. $\Delta V_{n-1,n}$ is the energy difference between CBM in adjacent layers, which shows the depinning between MoS$_2$ layers. $\Phi_B^e(n) = \text{CBM}_n - E_F$; $\Delta V_{n-1,n} = \text{CBM}_n - \text{CBM}_{n-1}$. (All values are given in eV)

| $n^{\text{th}}$ MoS$_2$ layer | $\Phi_B^e(n)$ | $\Delta V_{n-1,n}$ |
|:---:|:---:|:---:|
| 1$^{\text{st}}$ | 0.14 | -- |
| 2$^{\text{nd}}$ | 0.30 | 0.16 |
| 3$^{\text{rd}}$ | 0.42 | 0.12 |
| 4$^{\text{th}}$ | 0.52 | 0.10 |
| 5$^{\text{th}}$ | 0.56 | 0.04 |
| 6$^{\text{th}}$ | 0.57 | 0.03 |
| 7$^{\text{th}}$ | 0.58 | 0.01 |

As shown in Figure 3c, the charge density decays very fast (by an order of magnitude from the metal-MoS$_2$ interface to the MoS$_2$-MoS$_2$ interface), but $\Delta V_{n-1,n}$, which express the CBM difference between adjacent MoS$_2$ layers and represent the change in electrostatic potential, decays much slowly (Table I) than charge density does. The reason can be found by using the Poisson equation below. The $z$-direction MoS$_2$-layer-averaged charge density $\rho(z)$ and electrostatic potential $V(z)$ satisfy the one-dimensional Poisson equation [35]:

$$\nabla[\varepsilon(z)\nabla V(z)] = -\frac{\rho(z)}{\varepsilon} \qquad (4),$$

where $z$ is the distance from the electrode surface, $\varepsilon$ is the dielectric constant of MoS$_2$ and $\rho(z)$ is charge density, which includes the contributions from the energy level shift induced by charge transfer (or interface dipole), the energy level broadening (or IIGS) induced by interaction, and the MIGS, etc. Since $\rho(z)$ can be considered as the second-order derivative of $V(z)$, though the charge $\rho(z)$ decays rapidly with distance $z$ from the electrode surface, while the potential energy $-eV(z)$ decays much slowly and hence affects several MoS$_2$ layers, as shown in Figure 3b, 3d, and Table I.

As mentioned above, there's an *n*- to p-type contact transition in Pd and Pt junctions with MoS$_2$ layer-number change from monolayer to multilayer (Figure 2). Therefore, *p*-type contact is easier to be obtained in metal-*multilayered* MoS$_2$ junctions, which is useful for complementary-metal-oxide-semiconductor (CMOS) logic circuitry [41]. The transition is caused by both band gap decreases and the depining effect. It is well known that the band gap of $MoS_2$ decreases as the layer-number increases, and the biggest change occurs from monolayer to bilayer with bandgap transform from direct to indirect. In Figure 3a & 3b and Figure S5, CBM hardly changes with the increasing of $MoS_2$ layers since CBM is contributed from the first $MoS_2$ layer (with strong pinning) in metal-*multilayered* $MoS_2$ junction. In metal-*multilayered* $MoS_2$ junctions, the depinning effect of valence band is not so obvious (in Figure 1b) due to the indirect band character and band folding (see Figure S6).

## C. Universal route to type II band alignment

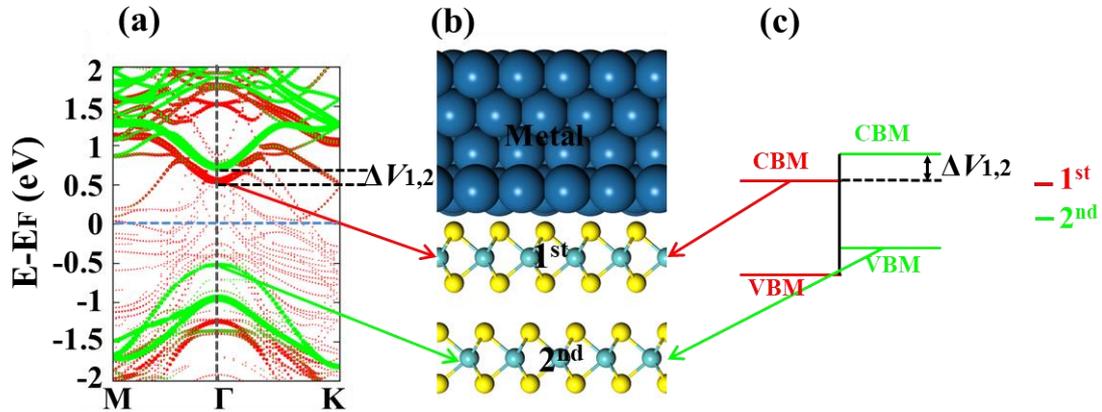

**FIG. 4. Type II band alignment between $MoS_2$ layers.** (a) Projected band structure to each layer of $MoS_2$ in Pt-2L $MoS_2$ junction. The first and second layers are shown by red and green dotted lines, respectively. (b) Geometric structure of Pt-2L $MoS_2$ junction. (c) Illustration of type II band alignment between $MoS_2$ layers, where red (green) indicates band edges from the first (second) layer.

According to the bands alignments of 2D semiconductor van der Waals junctions, semiconductor junctions can be classified into three types, straddling gap (type I), staggered gap (type II) and broken gap (type III) [42]. Due to the depinning between $MoS_2$ layers, type II band alignment is achieved in $MoS_2$ 'homojunctions' supported on metals, as shown in Figure 4. Taking 2L $MoS_2$ on Pt as an example, the CBM is contributed from the first $MoS_2$ layer and VBM is contributed from the second $MoS_2$ layer. It is worth noting that the band offset between the two $MoS_2$ layers is due to $\Delta V_{1,2}$, which is caused by the interface dipole pointing from the second $MoS_2$ layer to the first MoS$_2$ layer. This type II band alignment due to depinning between 2D semiconductor layers is universal in multilayered 2D semiconductors supported on metals and can apply straightforwardly to other 2D semiconductors supported on metals.

## 3. Conclusion

In summary, an extended Fermi-level pinning theory for M*m*SJ is proposed, the Schottky barrier and pinning factor *S* are both layer-number-dependent. The depinning between layers of vdW semiconductors in M*m*SJ results in the overall greatly weakened pinning, although strong pinning right at the metal–semiconductor interface. So, for the application to field effect transistors, Ohmic contact is relatively easy to obtain in M*m*SJ than in metal–monolayer 2D semiconductor junctions. Also due to depinning between 2D semiconductor layers, *p*-type contact, rarely obtained in 3D metal–monolayer 2D semiconductor junctions, becomes favored in MmSJ, which is useful for CMOS logic circuitry. The *p*-type contact with small/vanishing SBH is promising in M*m*SJ, by choosing large work-function metal and suitable multilayered 2D semiconductors [12]. Finally, the depinning between 2D semiconductor layers provide a new and universal route to type II band alignment (staggered band gap) for 2D semiconductor 'homojunctions' supported on metals, which is useful for optoelectronics due to the relatively easily obtained momentum space match [41] in homojunctions. Thus, our findings promote an emerging field, i.e. the layer-number engineering of 2D semiconductors in M*m*SJ for electronics/optoelectronics, where layer-number is a new degree of freedom for manipulating the pinning factor and SBH in MSJ, and for tuning the band-alignments in metal-supported 2D semiconductor 'homojunctions'.

## 4. Methods

Density-functional theory calculations are performed using the projector-augmented wave (PAW) [43] method as implemented in the VASP code [44,45]. A plane-wave cutoff energy of 500 eV was used to expand the valance electron wave functions. The surface Brillouin zone is sampled with a *k*-point grid spacing of 0.01 Å$^{-1}$ [46]. The geometric positions of the atoms are relaxed until the forces on each atom are smaller than 0.01 eV/Å, and the electronic optimization stops when the total energies difference of successive optimization loops smaller than $10^{-4}$ eV. We considered the influence of van der Waals (vdW) force by using optB88-vdW-DF [47-49], which offered a good performance compared with experimental results [50] for 2D semiconductors on metals.

The metal–multilayered MoS$_2$ junctions are composed of four layers of metal atoms (two outermost metal layers are fixed to preserve its bulk property) and 1L to 7L of MoS$_2$. Details about the different atomic relaxations of the different systems are shown in Supporting Information part II "structural relaxation in metal-MoS$_2$ junctions". A vacuum region of 15 Å is added to minimize the interaction between adjacent slabs, and a dipole correction is applied to avoid spurious interactions between periodic images of the slab [51]. The calculated MoS$_2$ in-plane lattice constant is a = 3.189 Å, which is in agreement with experimental and theoretical values [32,52]. During relaxation, the metal surface-lattices are strained to match MoS$_2$ (more discussions in Supporting Information) and the shape and size of the surface-cell is fixed with atoms fully relaxed [8]. Due to the very high computational cost for calculations containing 7L MoS$_2$, we select $\sqrt{3} \times \sqrt{3} R30°$ MoS$_2$ cells on top of $2 \times 2$ metal (111) surfaces for the five metal–MoS$_2$ junctions with 7L MoS$_2$, and in order to maintain the consistency, this small supercells are also used in junctions with layer-number of MoS$_2$. We demonstates that the size of supercells did not affect our conclusion (that the pinning factor S increase with the increasing MoS$_2$ layer number), more details are presented in part IV "effect of metal strain on pinning factor S" in Supporting Information.

## Acknowledgements

This work was supported by the Shenzhen Fundamental Research Foundation (Grant No. JCYJ20170817105007999), the Natural Science Foundation of Guangdong Province of China (Grant No. 2017A030310661), and the Guangdong Provincial Key Laboratory for Computational Science and Material Design (Grant No. 2019B030301001). The computational resources were provided by the Center for Computational Science and Engineering of Southern University of Science and Technology.

# Reference

[1] Jena D and Konar A 2007 *Phys. Rev. lett.* 98 136805

[2] Mak K F, Lee C, Hone J, Shan J and Heinz T F 2010 *Phys. Rev. lett.* 105 136805

[3] Dumcenco D, Ovchinnikov D, Marinov K, Lazić P, Gibertini M, Marzari N, Sanchez O L, Kung Y-C, Krasnozhon D, Chen M-W, Bertolazzi S, Gillet P, Fontcuberta i Morral A, Radenovic A and Kis A 2015 *ACS Nano* 9 4611-4620

[4] Gong C, Colombo L, Wallace R M and Cho K 2014 *Nano Lett.* 14,1714-1720

[5] Liu B, Wu L-J, Zhao Y-Q, Wang L –Z and Cai M-Q 2015 *Phys.Chem. Chem. Phys.* 17 27088-27093

[6] Yoon Y, Ganapathi K and Salahuddin S 2011 *Nano Lett.* 11, 3768-3773

[7] Allain A, Kang J, Banerjee K and Kis A 2015 *Nat. Mater.* 14 1195-1205

[8] Wang Q, Deng B and Shi X 2017 *Phys.Chem. Chem. Phys.* 19 26151-26157

[9] Jena D, Banerjee K and Xing G H 2014 *Nat. Mater.* 13 1076

[10] Zhao Y, Xu K, Pan F, Zhou C, Zhou F and Chai Y 2017 *Adv. Funct. Mater.* 27 1603484

[11] Liu Y, Stradins P and Wei S-H 2016 *Sci. Adv.* 2 e1600069

[12] Liu Y, Guo J, Zhu E, Liao L and Lee S J 2018 *Nature* 557 696

[13] Tung R T 2014 *Appl. Phys. Rev. 1* 4858400

[14] Schulman D S, Arnold A J and Das S 2018 *Chem. Soc. Rev.* 47 3037-3058

[15] Chen W, Santos E J G, Zhu W, Kaxiras E and Zhang Z 2013 *Nano Lett.* 13 509-514

[16] Kang J, Liu W, Sarkar D, Jena D and Banerjee K 2014 *Phys. Rev. X* 4 031005

[17] Kim C, Moon I, Lee D, Choi M S, Ahmed F, Nam S, Cho Y, Shin H-J, Park S and Yoo W J 2017 *ACS Nano* 11 1588-1596

[18] Popov I, Seifert G and Tománek D 2012 *Phys. Rev. Lett.* 108 156802

[19] Cui X, Lee G-H, Kim Y D, Arefe G, Huang P Y, Lee C-H, Chenet D A, Zhang X, Wang L, Ye F, Pizzocchero F, Jessen B S, Watanabe K, Taniguchi T, Muller D A, Low T, Kim P and Hone J 2015 *Nat. Nanotechnol.* 10 534-540

[20] Das S, Chen H-Y, Penumatcha A V and Appenzeller J 2013 *Nano Lett.* 13 100-105

[21] Jiang B, Zou X, Su J, Liang J, Wang J, Liu H, Feng L, Jiang C, Wang F, He J and Liao L 2018 *Adv. Funct. Mater.* 28 1801398

[22] Kim G-S, Kim S-H, Park J, Han K H, Kim J and Yu H-Y 2018 *ACS Nano* 12 6292-6300

[23] Kwon J, Lee J-Y, Yu Y-J, Lee C-H, Cui X, Honed J and Lee G-H 2017 *Nanoscale* 9 6151-6157

[24] Liu X, Qu D, Ryu J, Ahmed F, Yang Z, Lee D and Yoo W J 2016 *Adv.Mater.* 28 2345-2351

[25] Liu Y, Guo J, He Q, Wu H, Cheng H-C, Ding M, Shakir I, Gambin V, Huang Y and Duan X 2017 *Nano Lett.* 17 5495-5501

[26] Yu W J, Liu Y, Zhou H, Yin A, Li Z, Huang Y and Duan X 2013 *Nat. Nanotechnol.* 8 952-958

[27] Zhong H, Quhe R, Wang Y, Ni Z, Ye M, Song Z, Pan Y, Yang J, Yang L, Lei M, Shi J and Lu J 2016 *Sci. Rep.* 6 21786

[28] Zhong M, Xia Q, Pan L, Liu Y, Chen Y, Deng H-X, Li J and Wei Z 2018 *Adv. Funct. Mater.* 28 1802581

[29] Stengel M, Aguado-Puente P, Spaldin N A and Junquera J 2011 *Phys. Rev. B* 83 235112

[30] Shockley W 1939 *Phys. Rev.* 56 317-323

[31] Bardeen J 1947 *Phys. Rev.* 71 717-727

[32] Farmanbar M and Brocks G 2016 *Phys. Rev. B* 93 085304

[33] Peng H, Yang Z-H, Perdew J P and Sun J 2016 *Phys. Rev. X* 6 041005

[34] Hasegawa H and Sawada T 1983 *Thin Solid Films* 103 119-140

[35] Oehzelt M, Koch N and Heimel G 2014 *Nat. Commun.* 5 5174

[36] Ellis J K, Lucero M J and Scuseria G E 2011 *Appl. Phys. Lett.* 99 261908

[37] Shen T, Ren J-C, Liu X, Li S, and Liu W, 2019 *J. Am. Chem. Soc.* 141 3110

[38] Jin H, Li J, Wan L, Dai Y, Wei Y and Guo H 2017 *2D Mater.* 4 025116

[39] Otero R, Vazquez de Parga A L and Gallego J M 2017 *Surf. Sci. Rep.* 72 105

[40] Tung R T 2001 *Phys. Rev. B* 64 205310

[41] Taur Y, Ning T H, Fundamentals of Modern VLSI Devices, 2nd Ed, Cambridge University Press: Cambridge 2009 https://doi.org/10.1017/CBO9781139195065

[42] Ozcelik V O, Azadani J G, Yang C, Koester S J and Low T, 2016 *Phys. Rev. B* 94 035125

[43] Kresse G and Joubert D 1999 *Phys. Rev. B* 59 1758-1775

[44] Hafner J 2008 *J. Comput. Chem.* 29 2044-2078

[45] Kresse G and Furthmüller J 1996 *Phys. Rev. B* 54 11169-11186

[46] Monkhorst H J and Pack J D 1976 *Phys. Rev. B* 13 5188-5192

[47] Dion M, Rydberg H, Schröder E, Langreth D C and Lundqvist B I 2004 *Phys. Rev. Lett.* 92 246401

[48] Klimeš J, Bowler D R and Michaelides A 2011 *Phys. Rev. B* 83 195131

[49] Román-Pérez G and Soler J M 2009 *Phys. Rev. Lett.* 103 096102

[50] Liu W, Carrasco J, Santra B, Michaelides A, Scheffler M and Tkatchenko A 2012 *Phys. Rev. B* 86 245405

[51] J Neugebauer and Scheffler M 1992 *Phys. Rev. B* 46 16067

[52] Lin Z, Carvalho B R, Kahn E, Lv R, Rao R, Terrones H, Pimenta M A and Terrones M 2016 *2D Mater.* 3 022002